\documentclass[conference]{IEEEtran}
\IEEEoverridecommandlockouts

\usepackage{graphicx}
\usepackage{subcaption}
\usepackage{url}
\usepackage{hyperref}
\usepackage{amsmath}
\usepackage{booktabs}
\usepackage{multirow}
\usepackage{xcolor}
\usepackage{listings}
\usepackage{enumitem}
\usepackage{array}

\hypersetup{
    colorlinks=true,
    linkcolor=black,
    citecolor=black,
    urlcolor=blue
}

\title{DiSTILL: A Hybrid Cloud-HPC Workflow System for Reproducible Spatial Transcriptomics Analysis}
\author{Myles Joshua Toledo~Tan$^{\dagger}$,~\IEEEmembership{Member,~IEEE}\and Vasco Gerardo~Hinostroza Fuentes$^{\dagger}$,~\IEEEmembership{Student Member,~IEEE}\and Nikhil~Yerra \and Maria~Kapetanaki \and Parisa~Rashidi\and Kejun~Huang$^*$,~\IEEEmembership{Member,~IEEE} \and Panayiotis V.~Benos$^*$

\thanks{Myles Joshua Toledo~Tan is with the Department
of Electrical \& Computer Engineering, Herbert Wertheim College of Engineering, University of Florida, Gainesville, FL, USA, and also with the Department of Epidemiology, College of Public Health \& Health Professions and College of Medicine, University of Florida, Gainesville, FL, USA.}%
\thanks{Vasco Gerardo~Hinostroza Fuentes and Kejun~Huang are with the Department of Computer \& Information Science \& Engineering, Herbert Wertheim College of Engineering, University of Florida, Gainesville, FL, USA.}%
\thanks{Nikhil~Yerra is with the Clinical and Translational Science - Information Technology (CTS-IT) Team, Clinical and Translational Science Institute, University of Florida, Gainesville, FL, USA.}%
\thanks{Maria~Kapetanaki is with the Department of Pharmacotherapy \& Translational Research, College of Pharmacy, University of Florida, Gainesville, FL, USA.}%
\thanks{Parisa~Rashidi is with the J. Crayton Pruitt Family Department of Biomedical Engineering, Herbert Wertheim College of Engineering, University of Florida, Gainesville, FL, USA}%
\thanks{Panayiotis V.~Benos is with the Department of Epidemiology, College of Public Health \& Health Professions and College of Medicine, University of Florida, Gainesville, FL, USA.}%
\thanks{$^{\dagger}$These authors have contributed equally to this work and share first authorship.}
\thanks{$^*$Correspondence: Kejun~Huang, \url{kejun.huang@ufl.edu}; Panayiotis V.~Benos, \url{pbenos@ufl.edu}}
}

\begin{document}
\maketitle

\begin{abstract}
Spatial transcriptomics workflows increasingly combine large annotated data objects, notebook-based analyses, and resource-intensive statistical models that must be executed on high-performance computing (HPC) systems. In practice, these workflows are often difficult to reproduce because configuration, validation, stage execution, and artifact handling are fragmented across \textit{ad hoc} scripts and manually edited notebooks. We present \textit{DiSTILL} (Disease Diagnosis from Spatial Transcriptomics via Interpretable Latent Learning), a hybrid cloud--HPC workflow system for reproducible spatial transcriptomics (ST) analysis. DiSTILL combines an application programming interface (API) backend built with \texttt{FastAPI}, a web frontend, a dataset and preset registry, and a Python pipeline generator that materializes run-specific execution bundles and \texttt{SLURM} submission scripts. The system supports local, Secure Shell (SSH)-mediated, and pull-based poller execution modes, enabling HPC submission in environments where persistent API-initiated automation is restricted. We describe the system through the lens of an inflammatory bowel disease (IBD) ST workflow that operationalizes the analytical pipeline of Tan \textit{et al.} \cite{tan2025ibd} into an auditable application layer. Accordingly, the contribution of this paper is a workflow systems contribution centered on reproducible execution, queue-based orchestration, configuration semantics, and deployment across a split cloud--HPC architecture. The broader application goal of DiSTILL is to support user-supplied datasets that satisfy the schema assumptions of the wrapped analytical pipeline.
\end{abstract}


\section{Introduction}
Spatial transcriptomics (ST) analysis pipelines increasingly combine reference-guided deconvolution, latent factor discovery, notebook-based post-processing, and downstream predictive modeling. More broadly, the rapid growth of ST has produced a diverse computational landscape spanning preprocessing, spatial domain discovery, deconvolution, cell-cell interaction analysis, and multimodal or cross-slice integration~\cite{fang2023st,liu2024srtintegration,gaspardboulinc2025deconvolution}. Although many of the underlying analytical tools are well established, the software systems used to operationalize them remain fragile. In many research environments, execution still depends on manually edited paths, environment-specific shell commands, local notebook reruns, and one-off scheduler wrappers. These practices create failure modes that are infrastructural rather than methodological: hidden configuration drift, mismatched data schemas, incomplete provenance, duplicated queue execution, brittle file-path assumptions, and poor portability across compute environments~\cite{gruning2018reproducibility,gierend2024provenance,wilkinson2025fairworkflows}.

DiSTILL (Disease Diagnosis from Spatial Transcriptomics via Interpretable Latent Learning) was designed to address this systems gap. Rather than proposing a new spatial transcriptomics algorithm, DiSTILL provides an execution substrate that transforms preset-driven configurations into reproducible, auditable, and SLURM-ready runs. The platform separates configuration management, preflight validation, run materialization, queue handling, and artifact exposure into explicit software components. In this work, we present DiSTILL as a workflow systems contribution for computational biology.

As a case study, we deploy DiSTILL around the IBD ST pipeline introduced by Tan \textit{et al.}~\cite{tan2025ibd}, which used \texttt{cell2location}-derived cell-type abundance estimates, non-negative matrix factorization (NMF) latent representations (niches), niche composition features, neighborhood-enrichment features, and niche-associated gene expression features for machine learning--based disease classification~\cite{tan2025ibd}. DiSTILL does not replace that analytical contribution. Instead, it packages and orchestrates the workflow across a cloud-hosted control plane and an HPC execution environment.

A central design goal of DiSTILL is to move beyond a single hard-coded study and toward an application capable of processing user-supplied datasets, provided that they satisfy the structural and metadata assumptions of the wrapped analytical pipeline. In the current system, registry-backed datasets and preset-driven runs are implemented end to end, and upload-oriented flows are implemented for schema-constrained raw-entry and downstream dataset contracts. Cross-environment progress, log reflection, and artifact browsing are supported in deployment-sensitive ways through synced-artifact reporting from the HPC side. This paper therefore focuses on the software system architecture, operational semantics, and deployment trade-offs of DiSTILL.

\section{Related Work}
DiSTILL sits at the intersection of computational biology workflow systems and ST analysis.

\subsection{Analytical Tooling for Spatial Transcriptomics}
At the analytical level, DiSTILL integrates established computational tools. \texttt{Scanpy} provides scalable data structures and preprocessing utilities for large annotated single-cell and spatial gene expression matrices~\cite{wolf2018scanpy}, while the broader \texttt{scverse} ecosystem provides interoperable infrastructure for single-cell and spatial omics workflows~\cite{virshup2023scverse}. In particular, \texttt{AnnData} has become a common storage model for annotated matrices and associated metadata in these pipelines~\cite{virshup2024anndata}. \texttt{cell2location} provides a Bayesian framework for mapping reference-derived cell-type signatures into ST data~\cite{kleshchevnikov2022cell2location}. Additional spatial-analysis frameworks such as \texttt{Giotto}~\cite{dries2021giotto} and \texttt{Squidpy}~\cite{palla2022squidpy} support visualization, tissue composition analysis, neighborhood structure analysis, and other downstream spatial tasks. For the information-theoretic context surrounding the selection of niche-associated gene expression features and explainability, a recent review synthesizes how entropy, mutual information, and related criteria are used across ST analysis tasks~\cite{hinostroza2026infotheory}. The wrapped IBD workflow further incorporates NMF-based niche discovery, neighborhood enrichment analysis, feature engineering, and predictive modeling~\cite{tan2025ibd}. Hence, DiSTILL should be understood as an application and orchestration layer built around these existing analytical components.

\subsection{Workflow and Reproducibility Systems}
Workflow systems such as \texttt{Snakemake}~\cite{koster2012snakemake} and \texttt{Nextflow}~\cite{ditommaso2017nextflow} have established the importance of declarative workflows, reproducible execution, and scalable scheduling across local and cluster environments. \texttt{Galaxy} likewise emphasizes accessible, reproducible, and collaborative biomedical analyses through a web-facing workflow environment~\cite{afgan2018galaxy}. More recently, domain-oriented systems such as \texttt{Panpipes} and \texttt{SpatialOne} have automated substantial portions of multimodal single-cell and spatial transcriptomics analysis~\cite{curion2024panpipes,kamel2024spatialone}. These systems are valuable comparators, but translational ST workflows often require tighter integration with preset registries, dataset compatibility metadata, preflight validation, application-facing run creation, user progress views, and deployment in hybrid cloud--HPC environments where the control plane and execution environment are intentionally decoupled. DiSTILL addresses this gap by wrapping a research analysis workflow into an application layer with explicit run state, queue semantics, \texttt{SLURM} bundle generation, and artifact-aware execution.

DiSTILL is therefore positioned as an application-layer wrapper rather than as a replacement for general-purpose workflow engines. Its emphasis is not on introducing a new declarative execution language, but on operationalizing a concrete ST pipeline with preset-backed configuration, dataset compatibility checks, explicit preflight validation, user-facing run creation, and deployment-aware scheduler integration across a split cloud--HPC architecture.

\section{System Scope and Design Goals}
DiSTILL is intended as a reproducible workflow and deployment system for spatial transcriptomics pipelines. Its primary goals are to expose complex ST pipelines through a structured application interface rather than \textit{ad hoc} shell usage, preserve reproducibility by materializing every run as a concrete and inspectable execution bundle, validate configuration and selected data assumptions prior to expensive HPC submission, support multiple scheduler interaction patterns including settings where API-initiated unattended automation is constrained, and evolve toward user-supplied dataset processing under explicit schema and compatibility constraints.

These goals imply that DiSTILL should be evaluated as a workflow systems layer. The present paper therefore focuses on configuration semantics, queue handling, execution backends, deployment trade-offs, and operational lessons, while treating the wrapped IBD analysis pipeline as an integrated case study rather than as a new methodological contribution.

\section{Current Architecture}
DiSTILL is organized into four cooperating subsystems: a \texttt{Next.js} web frontend used for run setup, submission, and progress views; a \texttt{FastAPI} backend responsible for authentication, preflight validation, run creation, queue state, dataset and preset registry access, and artifact metadata; a Python pipeline generator that resolves a configuration into a concrete run directory containing \texttt{config.json}, \texttt{config.resolved.json}, a patched pipeline script, \texttt{run.sh}, and, when requested, \texttt{submit.sh}; and an HPC execution backend on HiPerGator (HPG) that executes materialized runs either through direct submission or through a pull-based poller that claims queued work from the API and submits it locally through \texttt{sbatch}.

The current production deployment is hybrid. The frontend is hosted separately from the API, while the API is served from a cloud virtual machine behind a reverse proxy. HPC execution occurs on the University of Florida HiPerGator environment through the Simple Linux Utility for Resource Management (\texttt{SLURM})~\cite{yoo2003slurm}. This separation is intentional: the control plane does not assume direct execution privileges on the cluster, and the pull-based poller avoids reliance on unattended API-initiated login-node automation.

\begin{figure*}[t]
    \centering
    \includegraphics[trim=0 1cm 0 0, clip, width=\textwidth]{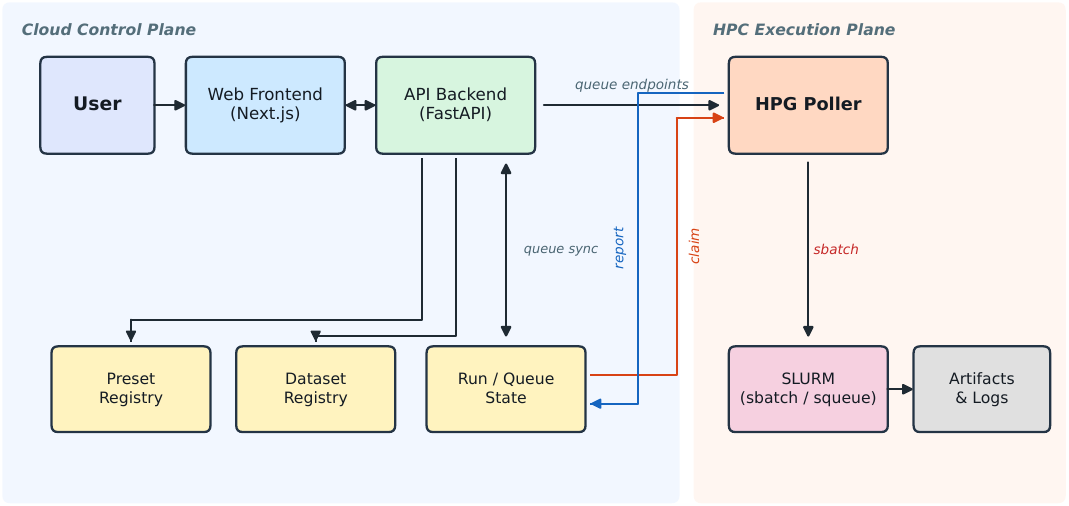}
    \caption{DiSTILL system architecture in the current deployment. The web frontend and API backend form a cloud-hosted control plane, while HiPerGator acts as the execution plane. Runs are created through the frontend and API, resolved against presets and dataset metadata, then exposed through authenticated queue endpoints. An HPG-resident poller claims queued runs, downloads prepared execution bundles, submits jobs locally via SLURM, and reports submission or status back to the API. This architecture avoids requiring a persistent API-initiated scheduler control channel into the HPC environment.}
    \label{fig:architecture}
\end{figure*}

\begin{table}[t]
\caption{DiSTILL components and current deployment roles.}
\label{tab:system_components}
\centering
\begin{tabular}{p{2.7cm} p{5.1cm}}
\toprule
\textbf{Component} & \textbf{Current role} \\
\midrule
Web frontend & \texttt{Next.js} user interface (UI) for run configuration, submission, and progress views \\
API backend & \texttt{FastAPI} control plane for auth, preflight, run creation, queue state, and artifact metadata \\
Pipeline generator & Resolves presets and dataset metadata into concrete run bundles and SLURM scripts \\
Preset registry & Stores workflow defaults, stage selection, and compute parameters \\
Dataset registry & Stores dataset-level compatibility metadata and schema hints \\
HPG poller & Claims queued work, materializes bundles on HPG, submits jobs via SLURM, and reports state \\
SLURM backend & Executes generated \texttt{submit.sh} and \texttt{run.sh} scripts on compute nodes \\
Artifacts/logs & Written to per-run output directories; visibility depends on storage accessibility and reporting mode \\
\bottomrule
\end{tabular}
\end{table}

Table~\ref{tab:system_components} complements Fig.~\ref{fig:architecture} by mapping each subsystem to its operational role in the current deployment.

\section{Execution Modes}
DiSTILL supports three operational backends built around the \texttt{SLURM} scheduler environment used in the current deployment~\cite{yoo2003slurm}.

\subsection{Local SLURM Backend}
In the local backend, the API host invokes \texttt{sbatch}, \texttt{squeue}, \texttt{sacct}, and related commands directly. This requires direct scheduler access and a shared filesystem view between the API host and the run and artifact paths.

\subsection{SSH-Mediated Backend}
In the SSH-mediated backend, the API host prepares a run locally, uploads the materialized directory to a remote system, and invokes SLURM commands through SSH. This supports remote scheduler access when the API host cannot run SLURM locally.

\subsection{Pull-Based HPC Poller}
In the pull-based mode, the API exposes authenticated queue endpoints while an HPC-resident poller periodically claims queued work, downloads a prepared bundle, materializes the run directory on HPG, submits the job locally, and reports submission or status back to the API. This pattern is especially useful when unattended API-initiated SSH automation is impractical or prohibited by HPC access policies.

\begin{figure}[t]
    \centering
    \includegraphics[width=\columnwidth]{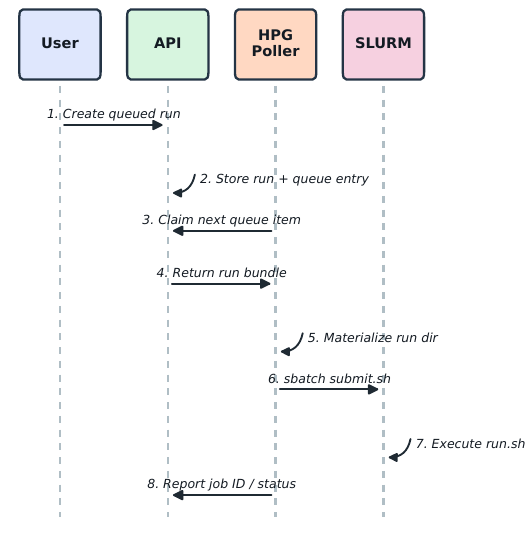}
    \caption{Execution sequence in pull-poller mode. The control plane creates a queued run and stores the resolved configuration. The HPG poller periodically claims available queue items using token-authenticated API calls, downloads the run bundle, materializes the run directory on HPG, submits the generated \texttt{submit.sh} script via local \texttt{sbatch}, and reports submission or subsequent state changes back to the API. Arrows indicating specific points mark the handoff point where controls move from the cloud control plane to the HPG execution side for local scheduler submission. Horizontal arrows denote message flow between actors, and the dashed vertical lifelines indicate actor persistence as time progresses from top to bottom.}
    \label{fig:poller_sequence}
\end{figure}

Figure~\ref{fig:poller_sequence} summarizes this pull-poller interaction pattern and the direction of control and status messages across the API, poller, and scheduler actors.

\subsection{Queue Lifecycle and Run-State Semantics}
DiSTILL separates logical run creation from physical job execution. This distinction is necessary because a run may be defined and validated before an execution resource is available, and in pull-poller mode the system that creates the run is not the same system that submits the job.

A run begins in a configuration-centric state such as \texttt{created} or \texttt{queued}. Once the API resolves the configuration and prepares an execution bundle, the run advances to a materialized state in which the generated scripts and resolved configuration are available for submission. In direct-submission modes, the API may immediately invoke \texttt{sbatch}. In pull-poller mode, the API instead exposes the run through authenticated queue endpoints and waits for an HPG-side poller to claim it. After successful claim and bundle download, the poller materializes the run on HPG, invokes \texttt{sbatch}, and reports the assigned SLURM job identifier back to the API. Subsequent states, such as \texttt{submitted}, \texttt{running}, \texttt{failed}, or \texttt{completed}, are updated from scheduler-facing information.

This lifecycle is intentionally decoupled from raw SLURM state names. The API must represent both application-layer state and scheduler-layer state, because a run can be validly prepared even when no SLURM job has yet been created. Likewise, a run may fail due to configuration or bundling issues before a scheduler identifier exists. Lease-based queue claiming further separates logical queue ownership from final scheduler submission and reduces the chance of duplicate processing by multiple pollers.

\section{Configuration and Data Model}
A run is defined by a JSON (JavaScript Object Notation) configuration that combines user-specified fields, preset defaults, and optional dataset-registry metadata. The configuration captures run metadata such as \texttt{run\_name} and selected stages, input data paths including spatial \texttt{AnnData} objects, reference \texttt{AnnData} objects, and metadata files, analysis parameters including NMF mode and stage-specific options, and execution parameters including \texttt{SLURM} resources and conda environment settings. This configuration strategy is consistent with the broader \texttt{scverse} ecosystem and the \texttt{AnnData}-centered data model commonly used for annotated single-cell and spatial matrices~\cite{virshup2023scverse,virshup2024anndata}.

Presets act as executable templates for recurring workflow families: they define stage order, default paths, parameter defaults, and scheduler resource profiles so that the same application surface can target different workflow variants without requiring script edits. The dataset registry stores compatibility hints, schema manifests, selected metadata fields, and recommended presets to reduce configuration drift and expose dataset-level assumptions at run-creation time. Separately, the API stores run metadata apart from queue metadata, allowing a run to exist in a valid application state before it acquires any scheduler identity; in pull-poller mode, queue entries additionally support lease-based locking.

\subsection{User-Supplied Dataset Processing}
A central design goal of DiSTILL is to support user-supplied datasets rather than only fixed, registry-backed studies. In practice, this means the application must either map uploaded datasets into the schema expected by the wrapped analytical pipeline or reject them during preflight validation with explicit feedback.

The intended scope should be stated carefully. DiSTILL is not designed to accept arbitrary unstructured files and automatically make them analyzable. Instead, it aims to accept user-supplied datasets that satisfy required structural and metadata assumptions. In the current deployment, registry-backed workflows are implemented end to end, and upload-oriented flows are available for explicit schema-constrained dataset contracts. For raw-entry workflows, the application expects a spatial \texttt{.h5ad} and matching metadata, while the required reference \texttt{.h5ad} may come either from the uploaded dataset record or from a compatible preset or registry-backed default. For downstream-entry workflows, the application can begin from an existing NMF-annotated artifact such as \texttt{cosmx\_with\_nmf.h5ad}. Files that do not satisfy required join-key, coordinate, morphology, or stage-input assumptions are surfaced during preflight rather than after \texttt{SLURM} submission. The operationally reliable path still benefits from controlled data placement, registry-backed defaults, and preflight validation, even though generalized upload support remains a longer-term product direction.

More generally, wrapping a new pipeline into DiSTILL requires only a small set of integration points that are explicit in the current architecture: a preset capturing stage selection and compute defaults, a dataset manifest or registry entry describing required paths and schema hints, a template script or notebook that can be materialized into a run-specific bundle, and stage-aware validation rules that define the expected inputs and outputs. This keeps the framework extensible without claiming arbitrary pipeline compatibility out of the box.

\section{Run Materialization}
DiSTILL does not execute directly from abstract configuration. Instead, each run is materialized into a concrete run directory containing \texttt{config.json}, \texttt{config.resolved.json}, a patched pipeline script, a stage orchestrator script (\texttt{run.sh}), a SLURM wrapper (\texttt{submit.sh}) when scheduler execution is enabled, and the generated artifacts and logs produced during execution.

This design preserves the exact code and configuration submitted to \texttt{SLURM} and supports \textit{post hoc} debugging because every run is represented as an inspectable filesystem object. The generated \texttt{run.sh} orchestrates stage execution sequentially, while \texttt{submit.sh} wraps the run for scheduler submission with environment activation, thread settings, and job-specific output and error targets.

\section{Preflight Validation}
Preflight validation reduces wasted queue time by checking configuration and selected data assumptions before execution. Depending on deployment mode, DiSTILL validates required fields and stage names, path existence and path safety relative to approved roots, dataset-registry compatibility constraints, and optional join-key compatibility between AnnData objects and metadata files. In HPC settings where scheduler time is expensive, surfacing these failures before submission is a substantial systems benefit.

\section{Case Study: Wrapping an IBD Spatial Transcriptomics Pipeline}
DiSTILL was developed around the IBD workflow by Tan \textit{et al.}~\cite{tan2025ibd}. Within DiSTILL, that workflow is treated as a set of orchestrated stages rather than as a novel method introduced by the platform. The wrapped stages may include \texttt{cell2location}-based spatial abundance inference~\cite{kleshchevnikov2022cell2location}, NMF-based niche decomposition, post-NMF neighborhood and enrichment analyses, \texttt{RCausalMGM}-oriented \cite{lovelace2026rcausalmgm} preparation steps, downstream predictive modeling including multilayer perceptron (MLP) analysis, and report generation, with optional SHapley Additive exPlanations (SHAP)-based feature attribution~\cite{lundberg2017shap}.

By wrapping these components in a system with explicit configuration, queueing, and artifact semantics, DiSTILL converts a research workflow into a more reproducible and application-oriented execution model.

\section{Case-Study Outputs From the Wrapped Pipeline}
\label{sec:results}
The following figures and tables provide a representative overview of the outputs generated by DiSTILL when hosting the wrapped IBD ST workflow. They are included as case-study evidence that the application layer can materialize a complete run, execute the expected stages, and expose representative artifacts spanning niche decomposition, neighborhood-enrichment features, classification summaries, statistical comparisons, and feature-importance outputs. They are not presented as a new biological contribution or as the sole evaluation of DiSTILL as a software system. For brevity, certain elements from the original analysis, e.g., causal graphs, have been omitted. For the complete biological results and comprehensive discussion, readers are referred to the original study~\cite{tan2025ibd}.

\setcounter{table}{2}

\subsection{Cellular Niche Decomposition}
\begin{figure*}[t]
    \centering
    \begin{subfigure}[b]{0.48\textwidth}
        \centering
        \includegraphics[width=\linewidth]{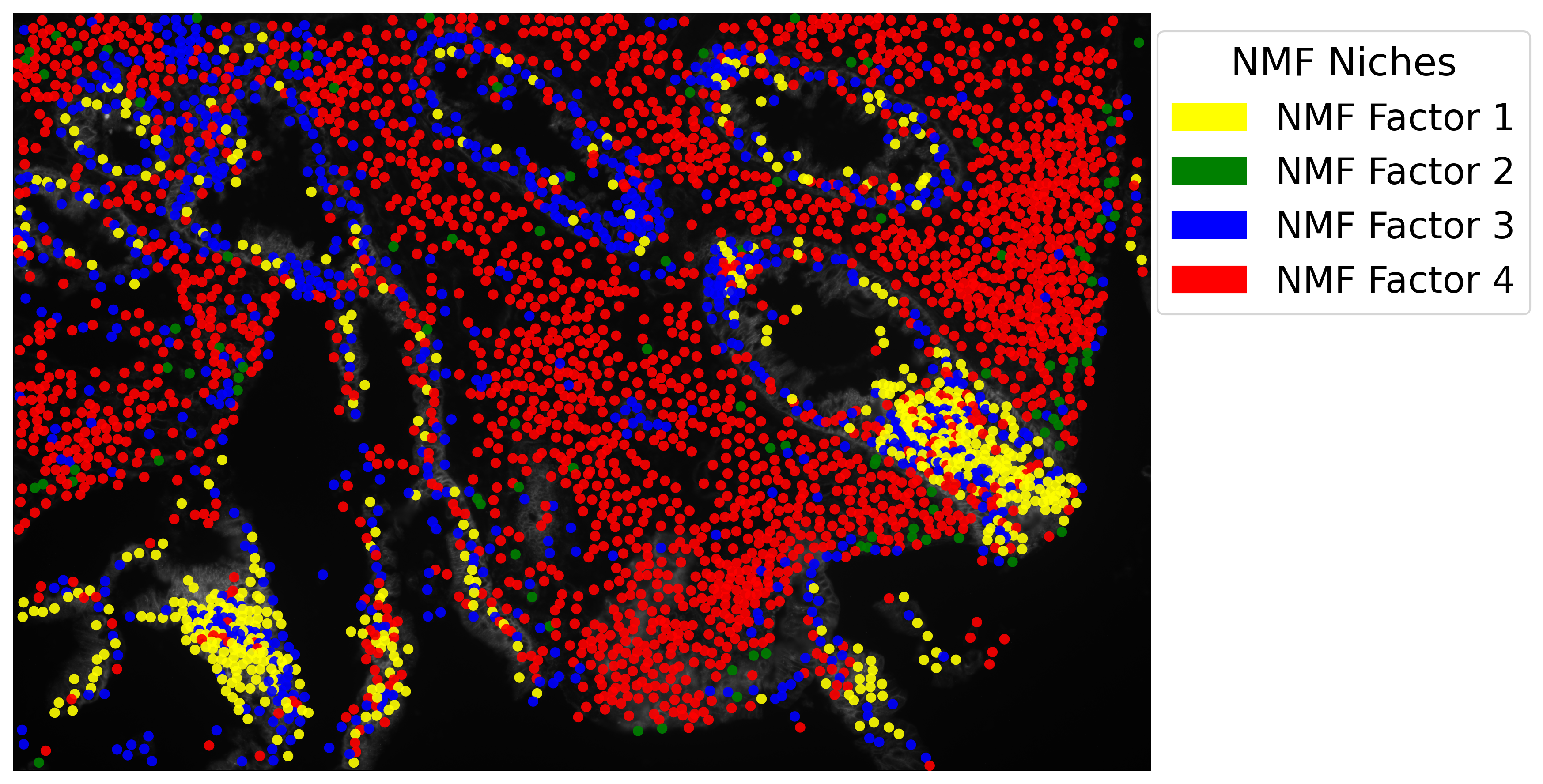}
        \caption{}
        \label{fig:nmf_mapping}
    \end{subfigure}
    \hfill
    \begin{subfigure}[b]{0.48\textwidth}
        \centering
        \includegraphics[width=\linewidth]{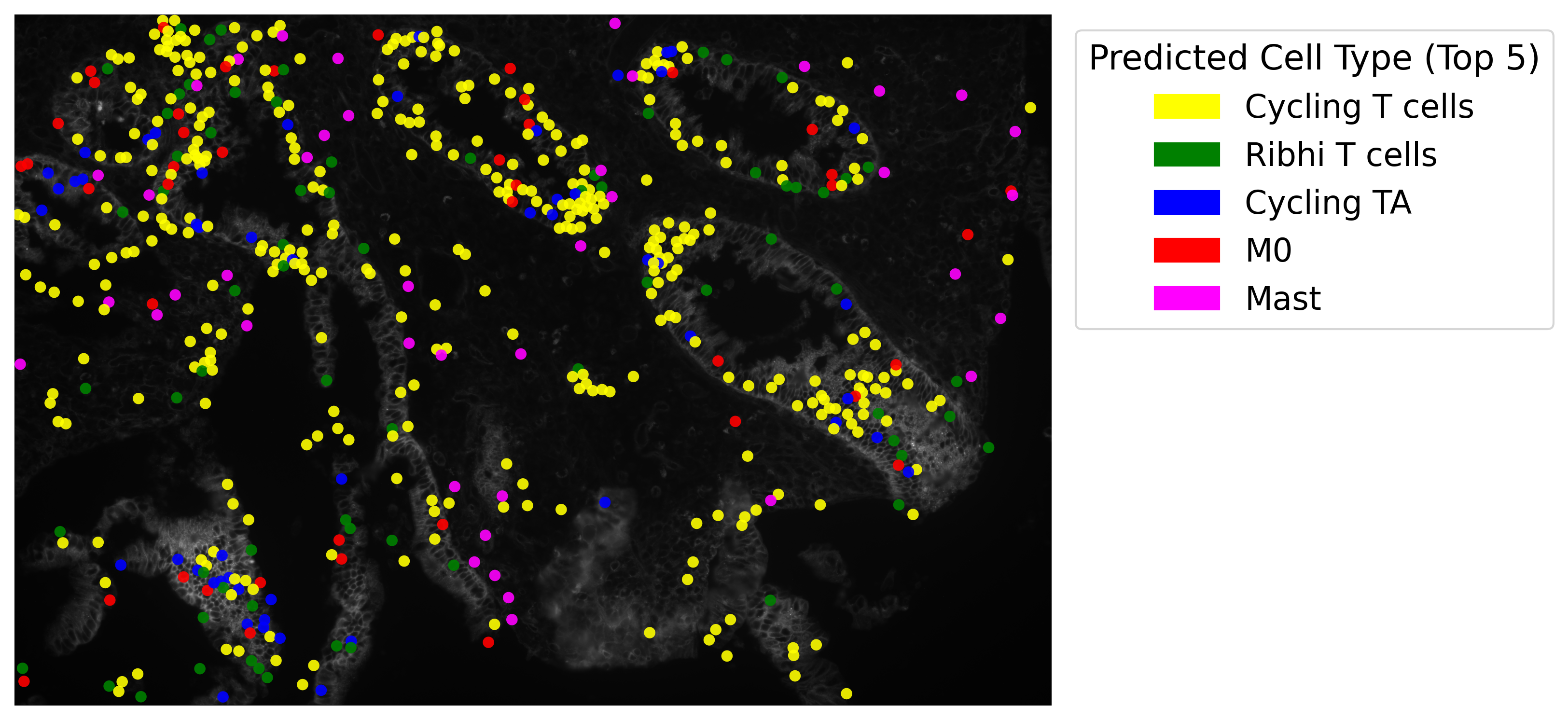}
        \caption{}
        \label{fig:top5_celltypes}
    \end{subfigure}

    \caption{Visualization of cellular niches and their cell type composition in FOV \texttt{UC\_a\_8}. (a) Cellular niches identified by NMF as colored points overlaid on the histomorphology. (b) The five most abundant cell types within Niche 3.}
    \label{fig:results_niches}
\end{figure*}

Figure~\ref{fig:results_niches} links computational niche discovery with biological interpretation in one representative field of view (FOV), \texttt{UC\_a\_8}. Panel (a) shows NMF-defined cellular niches overlaid on the histomorphology image, whereas panel (b) shows the five most abundant predicted cell types within Niche~3. This paired view makes the niche decomposition interpretable \textit{in situ} by connecting latent structure to tissue morphology and local cellular composition.

\subsection{Neighborhood Enrichment Features}
\begin{figure*}[t]
    \centering
    \begin{subfigure}[t]{0.28\textwidth}
        \centering
        \includegraphics[width=\linewidth]{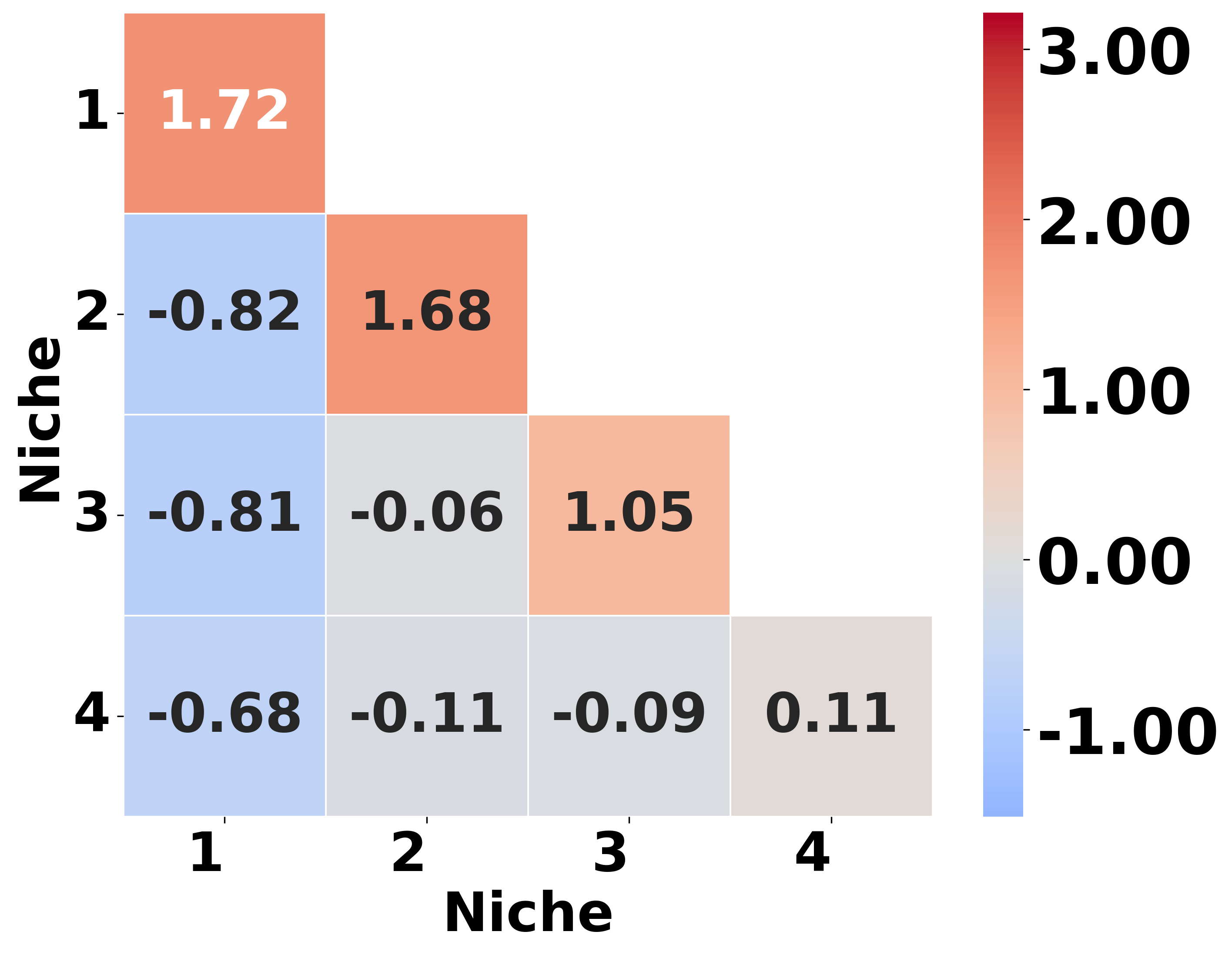}
        \caption{}
        \label{fig:hc}
    \end{subfigure}
    \hfill
    \begin{subfigure}[t]{0.28\textwidth}
        \centering
        \includegraphics[width=\linewidth]{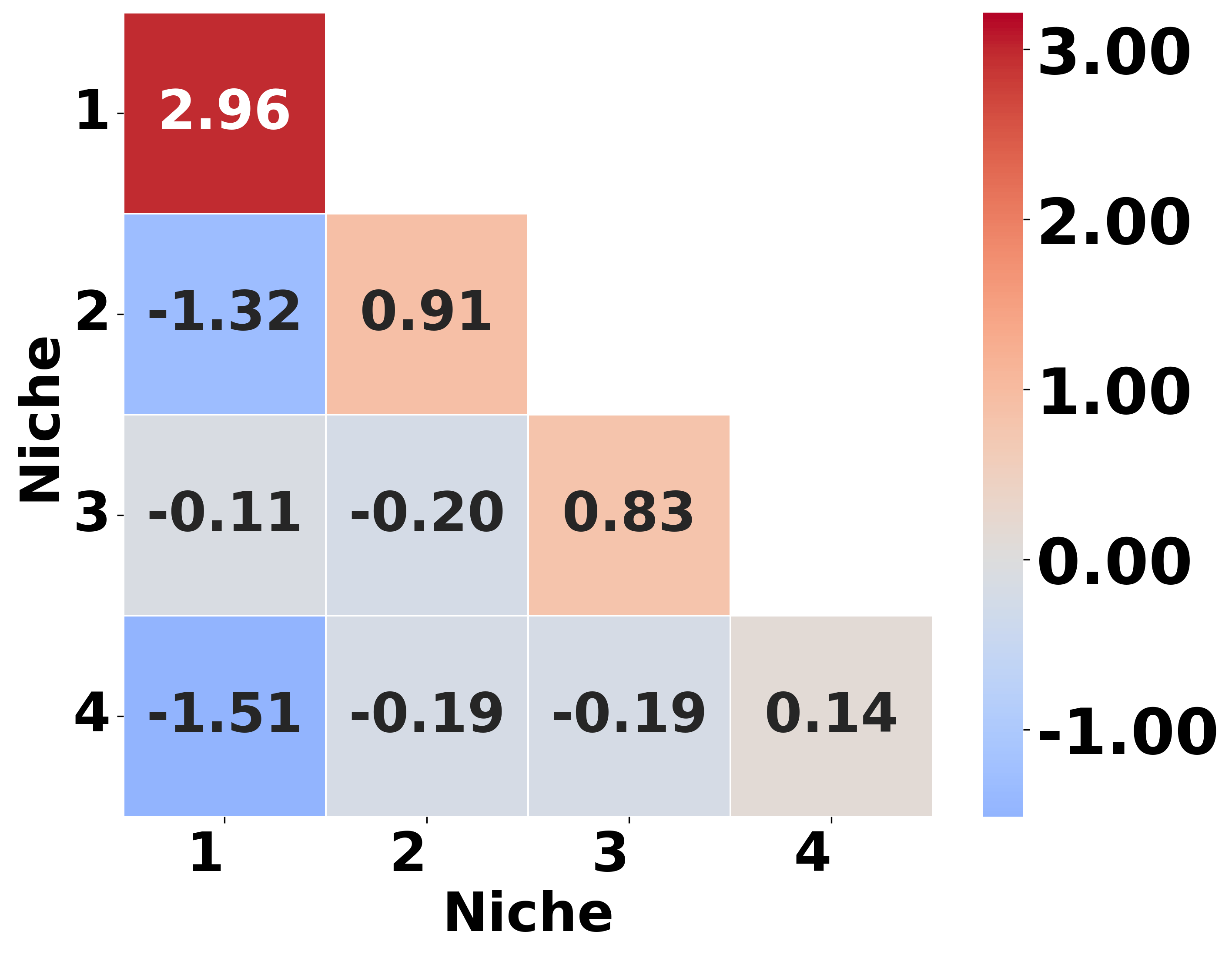}
        \caption{}
        \label{fig:uc}
    \end{subfigure}
    \hfill
    \begin{subfigure}[t]{0.28\textwidth}
        \centering
        \includegraphics[width=\linewidth]{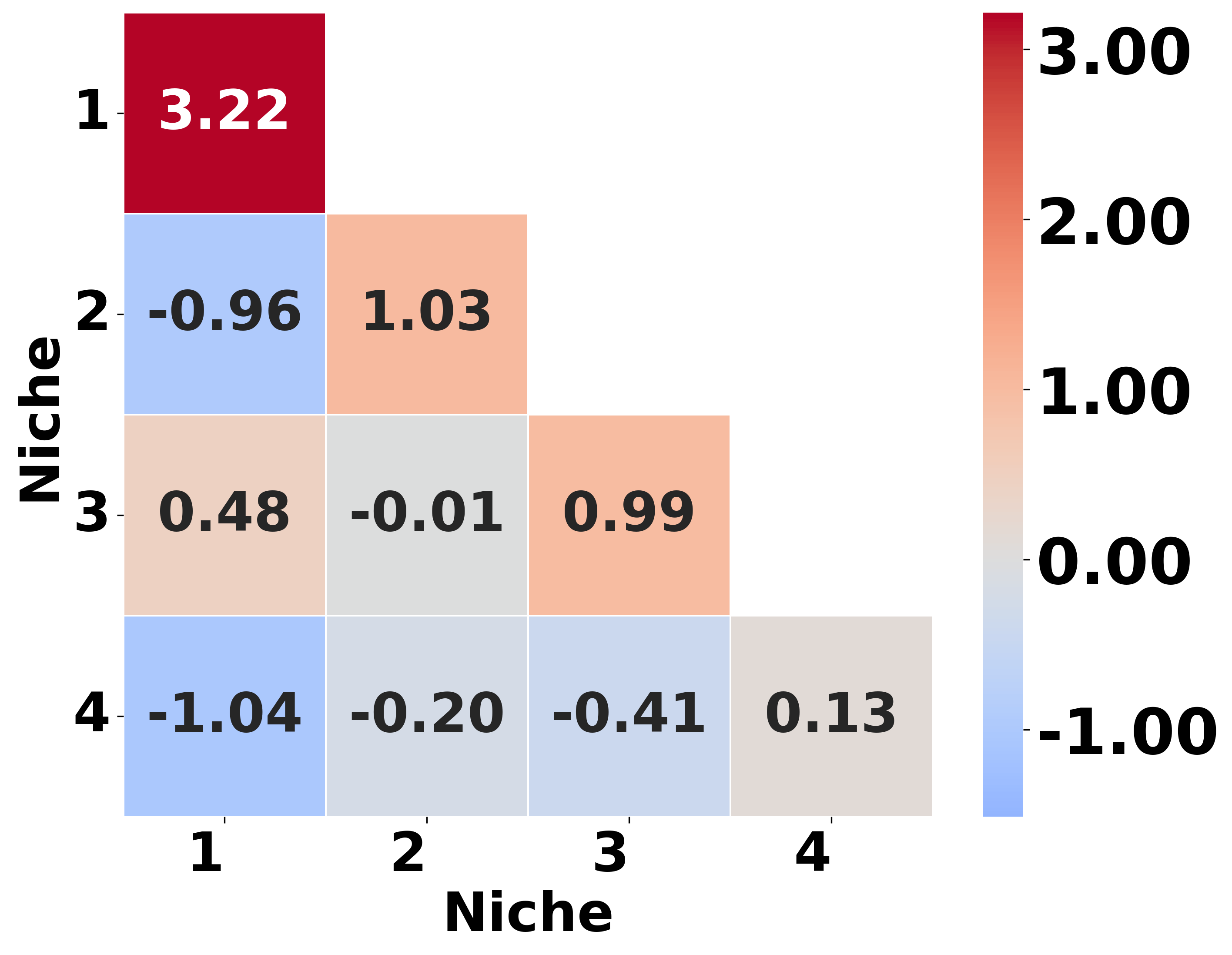}
        \caption{}
        \label{fig:cd}
    \end{subfigure}

    \caption{Niche enrichment comparisons across (a) HC, (b) UC, and (c) CD.}
    \label{fig:results_enrichment}
\end{figure*}

Figure~\ref{fig:results_enrichment} compares niche-neighborhood enrichment scores across HC, UC, and CD. Red cells denote niche pairs that co-occur more frequently than expected, whereas blue cells denote niche pairs that are under-enriched. The condition-specific contrast is especially useful for understanding how disease progression reshapes the spatial cellular microenvironment before downstream predictive modeling.

\subsection{Disease Classification Performance}
\begin{figure}[t]
\centering
\begin{subfigure}[b]{0.5\linewidth}
    \centering
    \includegraphics[width=\linewidth]{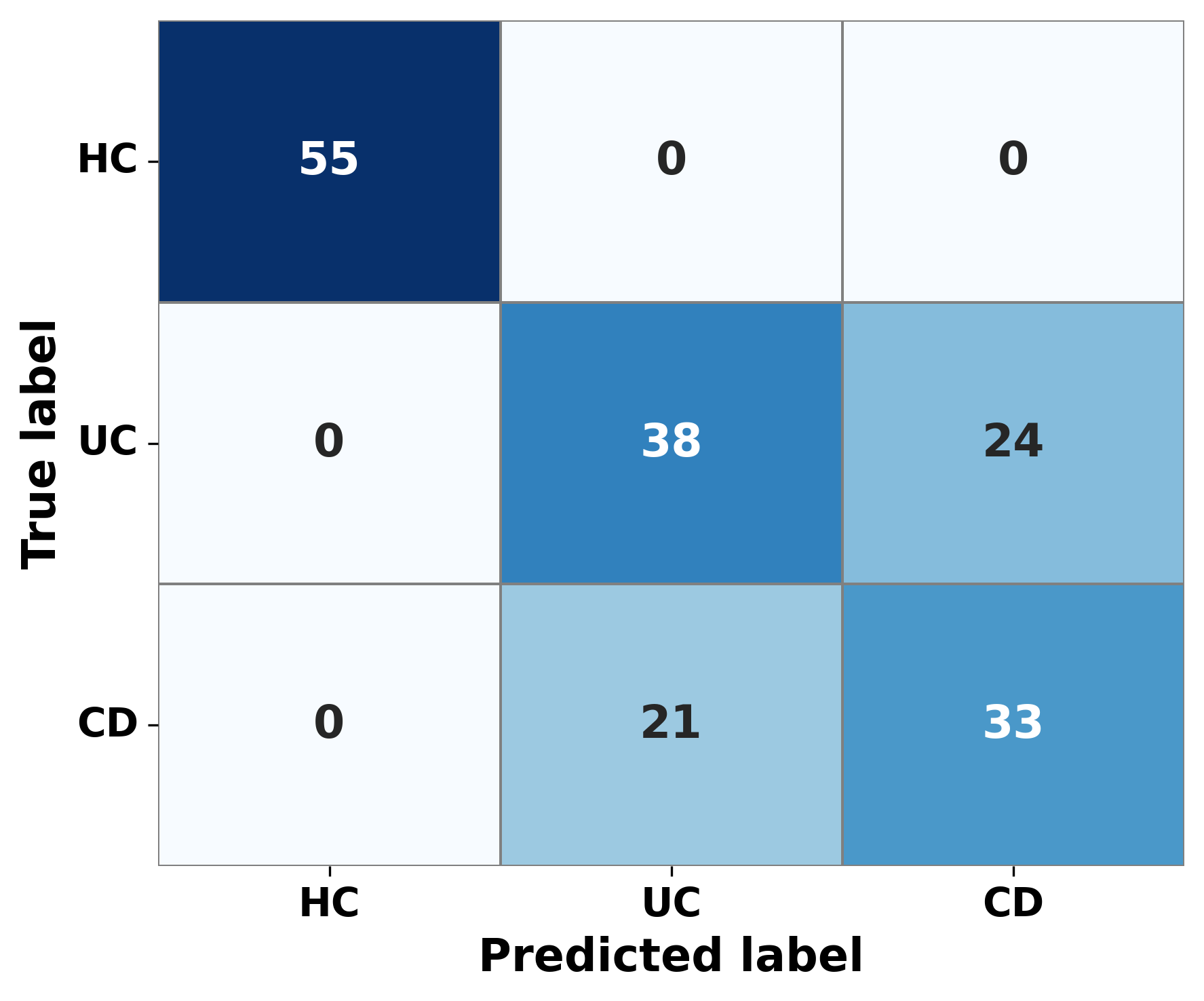}
    \caption{}
    \label{fig:cm_threeclass}
\end{subfigure}\hfill
\begin{subfigure}[b]{0.45\linewidth}
    \centering
    \includegraphics[width=\linewidth]{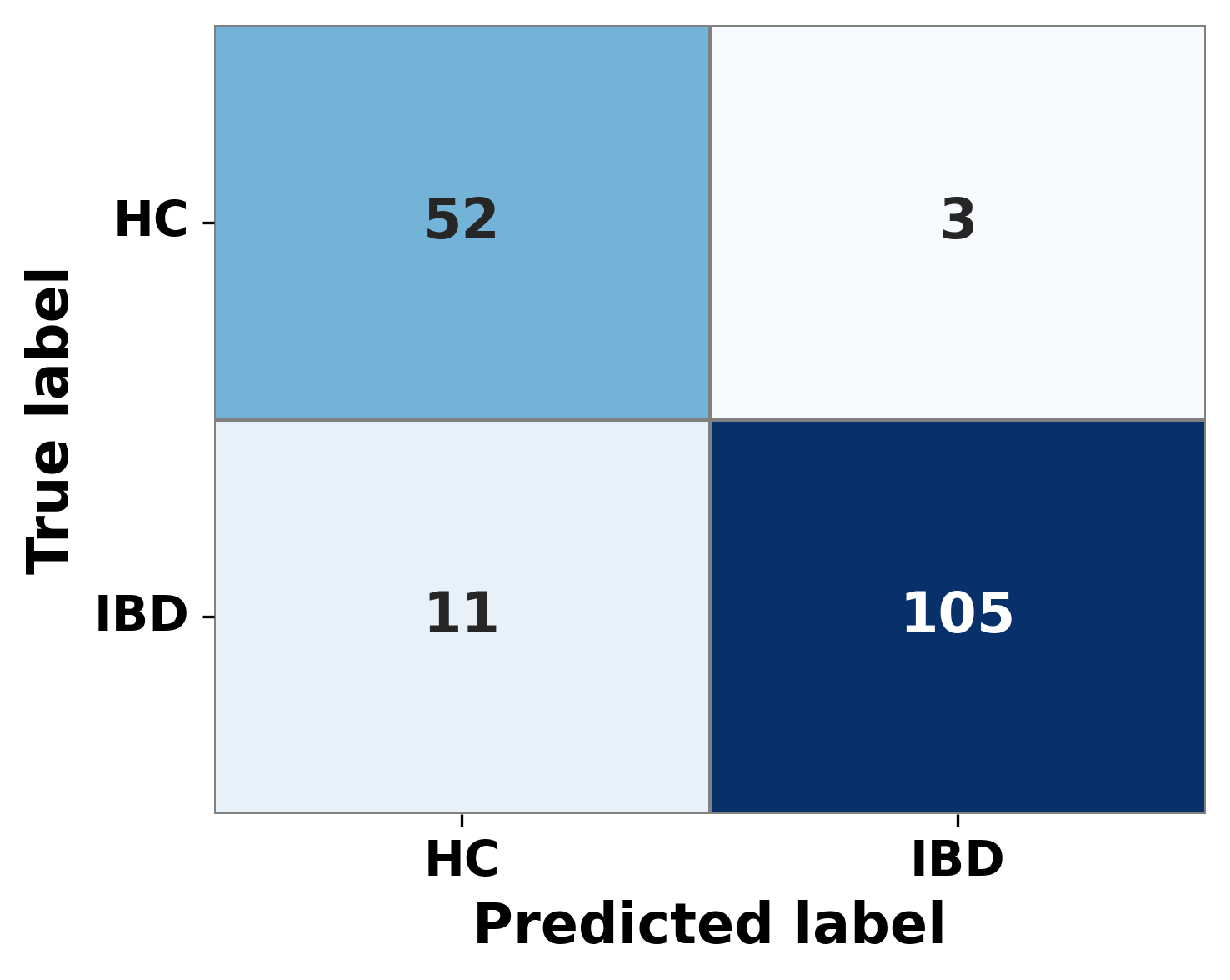}
    \caption{}
    \label{fig:cm_binary}
\end{subfigure}
\caption{Confusion matrices for the (a) three-class and (b) two-class problems.}
\label{fig:results_confusion}
\end{figure}

Figure~\ref{fig:results_confusion} summarizes the aggregated confusion matrices for the three-class and two-class MLP tasks. In the three-class setting, HC is classified perfectly, with most errors between UC and CD due to their biological similarity. Tables~\ref{tab:classification_three_class} and~\ref{tab:classification_two_class} provide the corresponding classification reports computed from the aggregated cross-validation outputs, and Table~\ref{tab:classification_cv} summarizes the mean and standard deviation of the evaluation metrics across the three folds.

\begin{table}[t]
\caption{Classification report for the three-class problem (HC, UC, CD).}
\label{tab:classification_three_class}
\centering
\small
\begin{tabular}{lcccc}
\toprule
\textbf{Class} & \textbf{Precision} & \textbf{Recall} & \textbf{F1-score} & \textbf{Support} \\
\midrule
HC & 1.000 & 1.000 & 1.000 & 55 \\
UC & 0.644 & 0.613 & 0.628 & 62 \\
CD & 0.579 & 0.611 & 0.595 & 54 \\
Accuracy & \multicolumn{3}{c}{0.737} & 171 \\
Macro Avg & 0.741 & 0.741 & 0.741 & 171 \\
Weighted Avg & 0.738 & 0.737 & 0.737 & 171 \\
\bottomrule
\end{tabular}
\normalsize
\end{table}

\begin{table}[t]
\caption{Classification report for the two-class problem (HC, IBD).}
\label{tab:classification_two_class}
\centering
\small
\begin{tabular}{lcccc}
\toprule
\textbf{Class} & \textbf{Precision} & \textbf{Recall} & \textbf{F1-score} & \textbf{Support} \\
\midrule
HC & 0.825 & 0.945 & 0.881 & 55 \\
IBD & 0.972 & 0.905 & 0.938 & 116 \\
Accuracy & \multicolumn{3}{c}{0.918} & 171 \\
Macro Avg & 0.899 & 0.925 & 0.909 & 171 \\
Weighted Avg & 0.925 & 0.918 & 0.919 & 171 \\
\bottomrule
\end{tabular}
\normalsize
\end{table}

\begin{table}[t]
\caption{Three-fold stratified group cross-validation performance for the three-class and two-class tasks.}
\label{tab:classification_cv}
\centering
\small
\begin{tabular}{lcc}
\toprule
\textbf{Performance metric} & \textbf{Three classes} & \textbf{Two classes} \\
 & \textbf{(HC, UC, CD)} & \textbf{(HC, IBD)} \\
\midrule
Accuracy & $0.774 \pm 0.161$ & $0.916 \pm 0.118$ \\
Precision & $0.743 \pm 0.220$ & $0.927 \pm 0.104$ \\
Recall & $0.741 \pm 0.183$ & $0.908 \pm 0.129$ \\
F1 Score & $0.712 \pm 0.209$ & $0.912 \pm 0.125$ \\
\bottomrule
\end{tabular}
\normalsize
\end{table}

\subsection{Statistical Comparisons and Feature Importance}
\begin{figure}[t]
\centering
\includegraphics[width=\columnwidth]{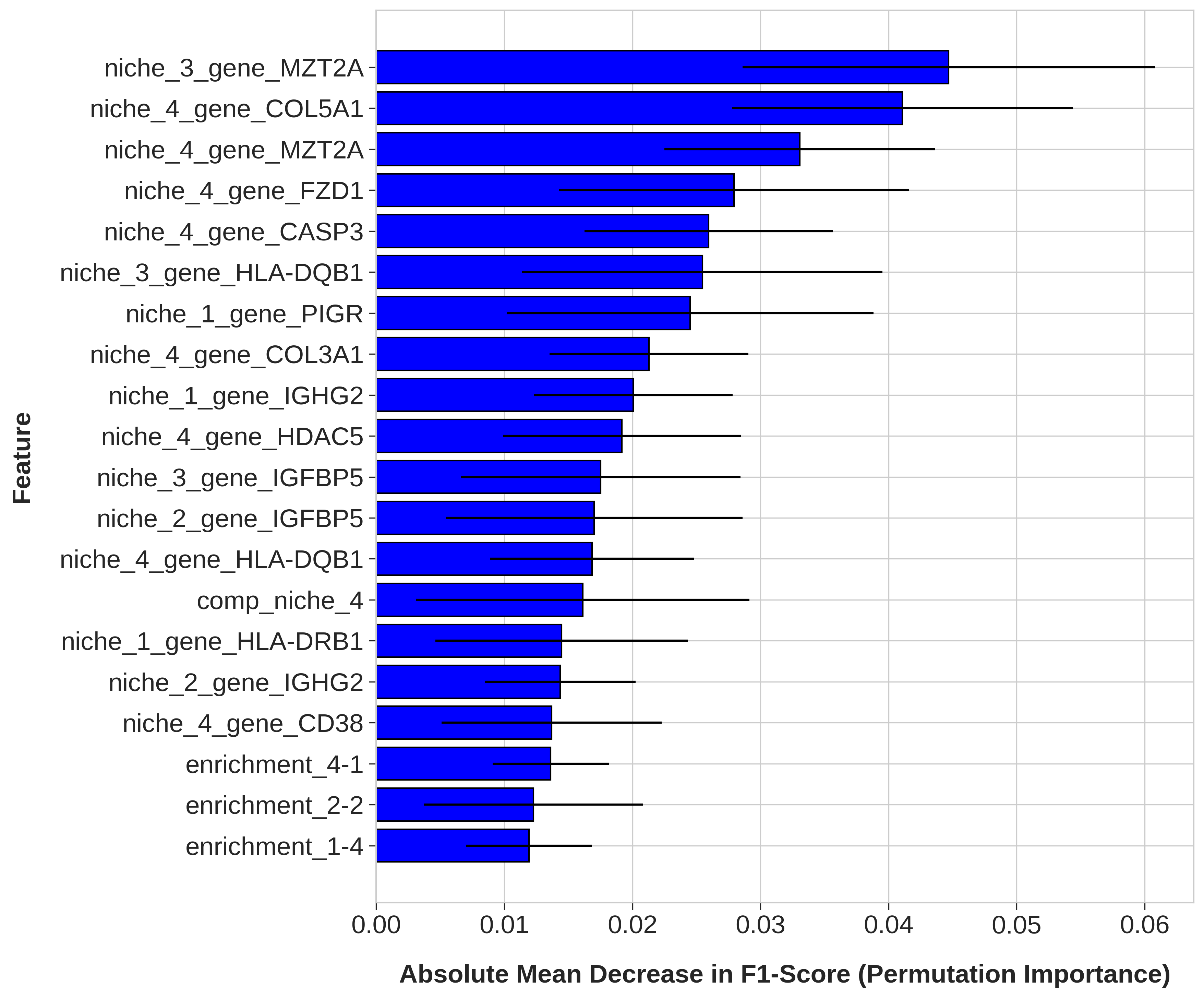}
\caption{Top 20 features ranked by PI for the three-class task.}
\label{fig:results_pi}
\end{figure}

Table~\ref{tab:niche_comparisons} reports the statistically significant pairwise comparisons for niche composition across disease groups, and Table~\ref{tab:interaction_comparisons} does the same for niche interactions. Together, these tables provide the direct inferential complement to the descriptive heatmaps in Fig.~\ref{fig:results_enrichment}. To preserve correspondence with the original study numbering, the final imported result figure is retained as Fig.~\ref{fig:results_pi}. Figure~\ref{fig:results_pi} shows the original permutation-importance analysis, highlighting the top 20 features for the three-class task.

\begin{table*}[t]
\caption{Pairwise comparisons of niche compositions across disease groups.}
\label{tab:niche_comparisons}
\centering
\small
\begin{tabular}{cccccl}
\toprule
\textbf{NMF factor} & \textbf{Group 1} & \textbf{Group 2} & \textbf{Difference between means} & \textbf{Direction} & \textbf{Adj.\ \textit{p}-value} \\
\midrule
1 & HC & UC & 7.2503 & Up & 0.0003 \\
1 & CD & HC & -10.2075 & Down & $< 0.0001$ \\
2 & HC & UC & -4.4167 & Down & $< 0.0001$ \\
2 & CD & HC & 2.4362 & Up & $< 0.0001$ \\
2 & CD & UC & -1.9805 & Down & 0.0219 \\
3 & HC & UC & -5.9405 & Down & $< 0.0001$ \\
3 & CD & UC & -7.3289 & Down & $< 0.0001$ \\
4 & CD & HC & 9.1598 & Up & $< 0.0001$ \\
4 & CD & UC & 12.2667 & Up & $< 0.0001$ \\
\bottomrule
\end{tabular}
\normalsize
\end{table*}

\begin{table*}[t]
\caption{Pairwise comparisons of niche interactions across disease groups.}
\label{tab:interaction_comparisons}
\centering
\small
\begin{tabular}{ccccccc}
\toprule
\textbf{Interaction} & \textbf{Group 1} & \textbf{Group 2} & \textbf{Direction} & \textbf{Group 1 mean} & \textbf{Group 2 mean} & \textbf{Adj.\ \textit{p}-value} \\
\midrule
1 vs 1 & HC & UC & Up & 0.165 & -2.627 & 0.032 \\
1 vs 1 & UC & CD & Down & -2.627 & -0.039 & 0.012 \\
1 vs 2 & HC & UC & Up & -1.308 & -3.652 & $< 0.001$ \\
1 vs 2 & HC & CD & Up & -1.308 & -2.650 & 0.014 \\
1 vs 4 & HC & UC & Up & -1.612 & -4.851 & 0.002 \\
2 vs 1 & HC & UC & Up & -1.693 & -3.673 & 0.001 \\
2 vs 1 & HC & CD & Up & -1.693 & -2.917 & 0.014 \\
4 vs 1 & HC & UC & Up & -1.736 & -4.823 & 0.002 \\
\bottomrule
\end{tabular}
\normalsize
\end{table*}

\section{System Evaluation}
Because DiSTILL is presented as a workflow systems contribution rather than as a new predictive model, the most relevant evaluation questions are operational: whether the system reduces manual execution burden, makes configuration failures more explicit, preserves reproducibility across reruns and recoveries, and improves the path from configuration to scheduler submission relative to \textit{ad hoc} execution.

\subsection{Operational Evidence from the Current Deployment}
The current implementation supports several concrete systems-level capabilities that are difficult to achieve reliably in an \textit{ad hoc} notebook-and-shell workflow. First, each run is materialized into a dedicated execution bundle containing a resolved configuration, patched pipeline script, stage orchestrator, and SLURM wrapper. This creates a stable submission artifact that can be inspected, rerun, or recovered after failure. Second, queue-based orchestration separates run definition from scheduler submission, which makes it possible to create a queued run from the control plane and later claim and submit it from HPG under institutional access constraints. Third, explicit poller callbacks for submission, status, and artifact synchronization provide a concrete bridge from the execution plane back to the control plane.

The case-study outputs summarized in Section~\ref{sec:results} provide functional evidence that the wrapped workflow can progress from niche decomposition through statistical testing and classifier interpretation within a reproducible application layer. This demonstrates that DiSTILL preserves enough state to support recovery-oriented execution and auditable result generation rather than forcing the user to reconstruct the workflow manually from notebooks and shell history.

\subsection{Comparison Axes Against Prior Ad Hoc Execution}
Table~\ref{tab:evaluation_axes} summarizes the most defensible evaluation axes for the current system. The comparison is intentionally qualitative because the present work does not include a controlled time-and-motion study or a formal user study. Nevertheless, these axes capture the concrete workflow differences between prior ad hoc execution and the current application layer.

\begin{table*}[t]
\caption{System-level evaluation axes comparing prior ad hoc execution with the current DiSTILL workflow.}
\label{tab:evaluation_axes}
\centering
\begin{tabular}{p{3.1cm} p{5.5cm} p{6.3cm}}
\toprule
\textbf{Evaluation axis} & \textbf{Prior ad hoc execution} & \textbf{Current DiSTILL workflow} \\
\midrule
Manual execution steps & Manual path edits, notebook reruns, direct shell invocation, hand-written SLURM commands, and manual log discovery across environments. & Preset-backed run creation, generated execution bundles, explicit stage orchestration via \texttt{run.sh}, generated \texttt{submit.sh}, and queue-driven HPG submission through the poller. \\
Configuration error exposure & Misconfigurations often surface only after scheduler submission or deep inside stage execution. & Preflight validation checks required fields, allowed roots, stage contracts, and selected dataset-schema assumptions before expensive execution. \\
Reproducibility across reruns & Reproduction depends on preserving shell history, notebook state, and manually edited file paths. & Each run is represented by concrete per-run artifacts such as \texttt{config.json}, \texttt{config.resolved.json}, patched pipeline code, and scheduler wrappers. \\
Time to submission & Submission requires environment-specific command knowledge and repeated manual setup on the execution side. & The control plane resolves and materializes the run bundle once; HPG submission then reduces to poller claim plus local \texttt{sbatch} invocation. \\
Failure recovery & Recovery often requires retracing prior commands and reconstructing which intermediate artifacts are valid. & Recovery can reuse preserved run artifacts, explicit stage boundaries, and synchronized outputs to rerun only the affected downstream stages. \\
Cross-environment visibility & Progress, logs, and outputs are fragmented across notebook outputs, shell sessions, and HPC filesystems. & Run state is reflected through the control plane, and selected artifacts and logs can be synchronized back from HPG for UI-facing access. \\
\bottomrule
\end{tabular}
\end{table*}

\subsection{What We Can and Cannot Claim}
The current evidence supports claims about operational structure, reproducible run representation, and deployment-aware workflow reliability, not a formal benchmark of human labor savings or scheduler efficiency. In particular, DiSTILL reduces the number of manual handoffs required to define, submit, and recover runs at the workflow-design level, but this paper does not claim a measured percentage reduction in configuration errors or a statistically validated improvement in wall-clock time to submission. A stronger future evaluation would include controlled before-versus-after measurements of operator effort, failure incidence, and recovery latency relative to the original ad hoc workflow, as well as broader user feedback across multiple deployments.

\section{Implementation Details}
The current implementation uses a \texttt{FastAPI} control plane, a \texttt{Next.js} web interface, filesystem-backed run directories, and \texttt{SQLite}-based run tracking. Queue safety uses lease-based claim semantics in pull-poller mode. Browser authentication is separated from queue-poller authentication, and artifact access is restricted to approved filesystem roots.

For HPC submission, DiSTILL generates job-specific SLURM output and error filenames to reduce collisions across repeated submissions. The generated submission wrapper standardizes conda activation and thread-related environment variables before invoking \texttt{run.sh}. In practice, this proved important during deployment because repeated submissions and large-memory jobs can make ambiguous shared log filenames difficult to debug.

\subsection{User Interface}
\begin{figure*}[t]
    \centering
    \includegraphics[width=0.94\textwidth,trim=10 10 10 10,clip]{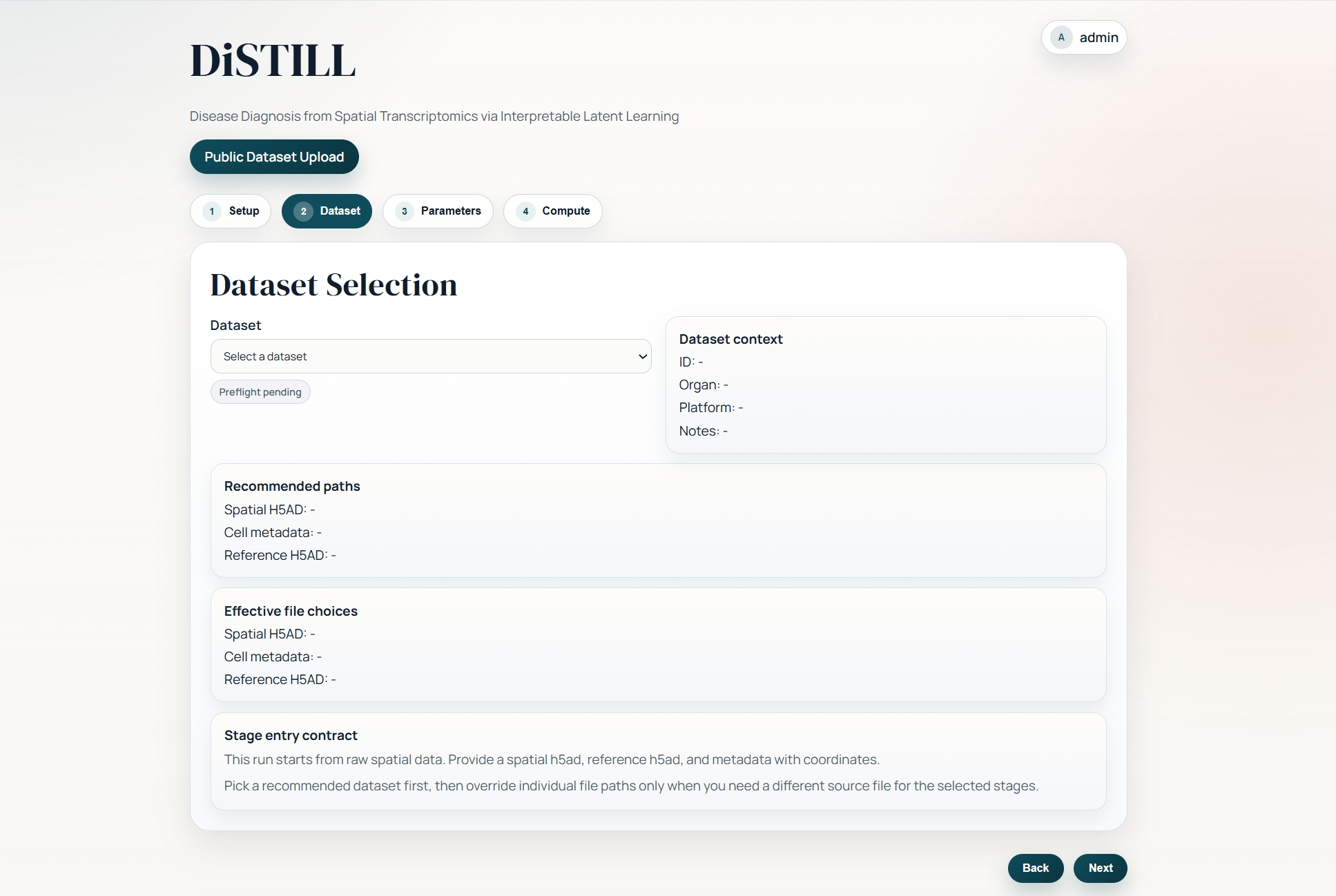}
    \caption{DiSTILL run-configuration user interface screenshot.}
    \label{fig:ui_run_setup}
\end{figure*}

The current manuscript includes representative user-interface captures so that the software paper explicitly documents the operator-facing surfaces in addition to the execution architecture. Figure~\ref{fig:ui_run_setup} shows the DiSTILL run-configuration view. DiSTILL also provides a separate progress-and-artifact view for tracking scheduler status, logs, and generated outputs.

\subsection{Implementation Notes}
Several implementation details were particularly important in making the system operational in a hybrid deployment. In practice, robustness depended on four choices: per-run materialization into inspectable execution directories; explicit separation between the cloud control plane and the HPG execution plane; token-authenticated poller access rather than browser-session automation for scheduler interaction; and layered logging that distinguishes scheduler-wrapper behavior from stage-level failures. Together, these choices simplify debugging, provenance tracking, reruns, and deployment across environments where the API host cannot directly control or inspect the cluster.

\section{Deployment and Reproducibility Considerations}
Artifacts are written under the configured run output directory, while stage logs are written by the stage orchestrator. In the split cloud--HPC deployment, user-facing visibility depends on whether HPG outputs are directly visible to the API host or are pushed back through a poller-mediated bridge into API-managed storage. This affects practical features such as live log reflection, artifact browsing, and public progress links, which must be generated against the frontend base URL rather than the API base URL.

To strengthen software-level reproducibility, the DiSTILL reference implementation is publicly available at \url{https://github.com/theocsav/DiSTILL}. The repository includes the \texttt{Next.js} frontend, \texttt{FastAPI} backend, pipeline runner, example presets, dataset-registry structure, and deployment documentation for local and HPC-oriented setups, as well as sanitized example configurations and smoke-test artifacts under \texttt{examples/}. It further documents supported dataset contracts and run/configuration interfaces in \texttt{docs/DATASET\_CONTRACT.md} and \texttt{docs/RUNSPEC\_RUNSTATE.md}. Because some datasets associated with the wrapped IBD case study may be subject to access or redistribution constraints, the public repository is intended to support instantiation and inspection of the DiSTILL system and its workflow contracts even when protected biological inputs cannot always be redistributed as a full reference deployment package.

\section{Deployment Lessons and Failure Modes}
Deploying DiSTILL across a cloud control plane and an HPC execution plane exposed several practical lessons that are relevant beyond this specific workflow.

A first lesson is that scheduler control assumptions must match institutional policy. In principle, an API host could invoke SLURM remotely through SSH. In practice, unattended login-node automation may be limited by authentication requirements or policy. The pull-poller design emerged as a robust alternative because it inverts the control flow: the HPC side retrieves work instead of the API pushing work inward.

A second lesson is that reproducibility failures often originate in operational glue rather than in the analytical core. During deployment and debugging, failures arose from issues such as mismatched hostnames for public progress views, environment activation differences between login and batch contexts, overwritten log files caused by non-unique filenames, scheduler wrapper assumptions, and latent script dependencies that were harmless in one environment but unavailable in another. These are not statistical failures, but they materially affect whether a workflow is usable.

A third lesson is that user-facing features become subtle in split deployments. A public progress page is a frontend concern, but scheduler state and logs originate from the API and HPC layers. If the frontend builds progress URLs against the API base URL rather than the frontend base URL, progress sharing fails even though the run state is valid. Similarly, live log browsing requires either shared storage visibility or a push-back mechanism from HPG to the API.

A fourth lesson concerns user-supplied datasets. Supporting uploads is not simply a matter of accepting a file through the frontend. The wrapped analytical pipeline makes assumptions about AnnData structure, metadata availability, and path layout. Consequently, uploads in the current system are implemented only for explicit schema-constrained contracts, such as raw-entry spatial-plus-metadata datasets and downstream NMF-annotated artifacts, and still benefit from registry-backed defaults and preflight validation. Upload workflows are therefore supported, but they are not arbitrary opaque file ingestion.

These lessons reinforce the main argument of the paper: the gap between a research script collection and a reproducible application is primarily a software systems problem.

\section{Limitations}
Several limitations should be stated explicitly. The current system is oriented toward single-instance deployment, and SQLite limits multi-writer scalability. In pull-poller deployments, artifact and live-log visibility depend on either shared storage visibility or explicit push-back from HPG to the API, and synced reflection remains configuration-sensitive. Public progress links and log surfacing also remain deployment-sensitive across the split frontend/API architecture. In addition, the user-dataset upload path is limited to explicit schema-constrained contracts rather than unrestricted ingestion of arbitrary files, and some wrapped analytical scripts still require dependency and path hardening for full portability across arbitrary datasets and execution environments. Finally, while the present qualitative evaluation is appropriate to the current systems contribution, broader quantitative study across additional users and deployments remains future work.

\section{Discussion}
DiSTILL demonstrates that turning a research-grade spatial transcriptomics workflow into an application requires solving systems problems that are often omitted from methodological papers: configuration normalization, path safety, scheduler interaction, queue duplication prevention, reproducible bundle generation, run provenance, and deployment-aware artifact access. A useful interpretation of DiSTILL is therefore as a translational layer between computational biology research code and application-facing workflow execution, making assumptions explicit, separating validation from execution, and representing each run as a concrete auditable object.

\section{Conclusion}
DiSTILL is a reproducible workflow orchestration system for spatial transcriptomics analysis in hybrid cloud--HPC environments. Its contribution is not a new biological model, but a systems layer that operationalizes complex computational biology pipelines through configuration resolution, preflight validation, run materialization, queue semantics, execution backends, and artifact traceability. By wrapping the IBD ST pipeline of Tan \textit{et al.} into an application framework and generalizing toward user-supplied datasets, DiSTILL demonstrates how research analysis code can be elevated into a more reproducible and usable workflow system.

\section*{Acknowledgment}
This work was supported by the National Institutes of Health (NIH) grants. ChatGPT (GPT-5.4) was used solely to optimize phrasing and reduce wordiness in order to meet the page limits set by the conference. All ideas, analyses, and conclusions presented in this manuscript are entirely our own.

\bibliographystyle{IEEEtran}
\bibliography{references}

\end{document}